\documentclass[a4paper,11pt]{article}
\pdfoutput=1
\usepackage{jheppub}
\usepackage{breakurl}
\usepackage{mathrsfs}
\usepackage{amsmath}
\usepackage{amsfonts}
\usepackage{mathrsfs}
\usepackage{slashed}
\usepackage{epigraph}
\usepackage{fancyhdr}
\usepackage{amssymb}
\usepackage{graphicx}
\usepackage{longtable}
\usepackage{makecell}
\usepackage{amscd}
\usepackage{tabu}
\usepackage{bm}
\usepackage{color}
\usepackage{hyperref} 
\hypersetup{linktocpage=true}
\usepackage[all]{hypcap}
\hypersetup{
   bookmarks=true,         
   unicode=false,          
   pdftoolbar=true,        
   pdfmenubar=true,        
   pdffitwindow=false,     
   pdfstartview={FitH},    
   pdftitle={My title},    
   pdfauthor={Author},     
   pdfsubject={Subject},   
   pdfcreator={Creator},   
   pdfproducer={Producer}, 
   pdfkeywords={keyword1} {key2} {key3}, 
   pdfnewwindow=true,      
   colorlinks=true,        
   linkcolor=blue,         
   citecolor=magenta,      
   filecolor=magenta,      
   urlcolor=cyan           
}
\usepackage{fancyhdr}

\def\beq{\begin{equation}} 
\def\eeq{\end{equation}} 
\def\bea{\begin{eqnarray}} 
\def\eea{\end{eqnarray}} 
\def\nn{\nonumber}

\def\bsll{ b \to  s \, \ell^+ \, \ell^-}
\def\Bsmumu{B_s \to \mu^+ \mu^-}
\def\Bsee{B_s \to e^+ e^-}
\def\Bsll{B_s \to \ell^+ \ell^-}
\def\BXsmumu{B \to X_s \mu^+ \mu^-}
\def\BXsee{B \to X_s e^+ e^-}
\def\BXsll{B \to X_s \ell^+ \ell^-}
\def\BKmumu{B^+ \to K^+ \mu^+ \mu^-}
\def\BKee{B^+ \to K^+ e^+ e^-}
\def\BKll{B^+ \to K^+ \ell^+ \ell^-}
\def\BKstarmumu{B \to K^* \mu^+ \mu^-}
\def\BKstarll{B \to K^* \ell^+ \ell^-}

\title{Hint of Lepton Flavour Non-Universality in $B$ Meson Decays} 
\author[a,1]{Diptimoy Ghosh}
\author[b,2]{Marco Nardecchia}
\author[b,3]{S. A. Renner}

\affiliation[a]{\normalfont{INFN, Sezione di Roma, Piazzale A. Moro 2, 
I-00185 Roma, Italy}}
\affiliation[b]{\normalfont{DAMTP, University of Cambridge, Wilberforce Road, 
Cambridge CB3 0WA, United Kingdom}}

\emailAdd{diptimoy.ghosh@roma1.infn.it} 
\emailAdd{m.nardecchia@damtp.cam.ac.uk}
\emailAdd{sar67@cam.ac.uk} 

\abstract{
The LHCb collaboration has recently presented their result on \\
$R_K =\mathcal{B} \left( B^+ \to K^+ \mu^+ \mu^-\right)/ \mathcal{B} \left( B^+ \to 
K^+ e^+ e^-\right)$ for the dilepton invariant mass bin $m_{\ell \, 
\ell}^2 = 1-6$ GeV$^2$ ($\ell = \mu, e$). The measurement shows an 
intriguing $2.6 \, \sigma$ deviation from the Standard Model (SM) prediction. 
In view of this, we study model independent New Physics (NP)  
explanations of $R_K$ consistent with other measurements involving $\bsll$ 
transition, relaxing the assumption of lepton universality. We perform a 
Bayesian statistical fit to the NP Wilson Coefficients and compare the Bayes 
Factors of the different hypotheses in order to quantify their goodness-of-fit. We 
show that the data slightly favours NP in the muon sector over NP in the electron 
sector. 
}
\begin{document} 
\maketitle
\flushbottom
\section{Introduction}

While we all hope to see direct evidence of new particles in the upcoming 14 
TeV run of the LHC, indirect searches for New Physics (NP) through precision 
measurements are also extremely important, in particular because of their 
high sensitivity to Ultra Violet (UV) physics. In fact, the 8 TeV run of the 
LHC has already seen interesting indirect hints of NP in some 
of the $B$ meson decay modes. The quoted 3.7 $\sigma$ deviation observed by 
the LHCb collaboration last summer \cite{Aaij:2013iag,Aaij:2013qta} in one of 
the angular observables ($P_5^\prime$) in the decay $\BKstarmumu$ for one of 
the dilepton invariant mass bins ($4.30 < q^2 = m_{\mu \, \mu}^2 < 8.68$ 
GeV$^2$) inspired many theorists to come up with NP explanations 
\cite{Descotes-Genon:2013wba,Altmannshofer:2013foa,Gauld:2013qba,Gauld:2013qja, 
Datta:2013kja,Buras:2013qja,Buras:2013dea,Altmannshofer:2014cfa}. Interestingly enough, very 
recently the LHCb collaboration has observed a 2.6$\sigma$ deviation 
in another $\bsll$ mode; in the quantity called $R_K$ which is the ratio of 
the two branching fractions $\mathcal{B} \left( B^+ \to K^+ \mu^+ 
\mu^-\right)$ and $\mathcal{B} \left( B^+ \to K^+ e^+ e^-\right)$ in the 
dileptonic invariant mass bin \mbox{$m_{\ell \, \ell}^2 = 1-6$ GeV$^2$} 
($\ell = \mu, e$) \cite{Aaij:2014ora}. Note that the branching ratios 
$\mathcal{B} \left( B^+ \to K^+ \mu^+ \mu^-\right)$ and $\mathcal{B} \left( 
B^+ \to K^+ e^+ e^-\right)$ are individually predicted with very large 
hadronic uncertainties ($\sim 30\%$) in the SM \cite{Bobeth:2007dw}. However, 
their ratio is a theoretically very clean observable and predicted to be 
$R_K^{\rm SM} = 1$ if lepton masses are ignored \cite{Hiller:2003js}. 
Inclusion of the lepton mass 
effects changes the prediction only by a tiny amount making it $R_K^{\rm SM} 
= 1.0003 \pm 0.0001$ \cite{Bobeth:2007dw}. This is in contrast to the 
($P_5^\prime$) anomaly where considerable debate exists surrounding the issue 
of theoretical uncertainty due to (unknown) power corrections to the 
factorization framework 
\cite{Jager:2012uw,Lyon:2014hpa,Descotes-Genon:2014uoa}. Hence, there is a 
possibility that the observed deviation might be (partly) resolved once these 
corrections are better understood. On the other hand, the observable $R_K$ is 
a ratio of two branching fractions which differ only in the flavour of the final state leptons.  
This makes $R_K$ well protected from hadronic uncertainties in the SM because the strength of the gauge interactions contributing to the short distance physics in the $\bsll$ 
transition are independent of the final state lepton flavour. In fact, this 
feature remains true even in NP models if the model respects lepton 
universality. Therefore, the $R_K$ measurement is perhaps pointing towards 
new short distance physics which is not lepton universal. This motivated us 
to study possible NP explanations of the $R_K$ measurement (and their 
consistency with other observables involving the $\bsll$ transition) in a 
model independent way, relaxing the assumption of lepton universality. To 
this end, we perform a statistical fit to the NP Wilson Coefficients (WC) 
employing Bayesian inference. In order to quantify and compare the 
goodness-of-fit of the different hypotheses we also compute their relative 
Bayes Factors (BFs). 

The paper is organized as follows. In the next section we set up our notation 
and discuss all the experimental data used in our analysis. We present our 
results in section \ref{results} and conclude with some final remarks in 
section \ref{conclusion}.

\section{Effective Field Theory approach}
\label{eft}

We base our analysis on the following 
$\left| \Delta B \right| = \left| \Delta S  \right| = 1$ effective 
Hamiltonian,
\beq
{\mathcal H}_{\rm eff} = - 
\frac{4 G_F}{\sqrt{2}}\, (V_{ts}^\ast V_{tb}) \,  
\sum_{i}^{} \hat{C}^{\ell}_i (\mu) \, {\mathcal O}^{\ell}_i (\mu) \, \, ,
\eeq
where $\mathcal{O}^{\ell}_i$ are the $\rm SU(3)_C \times U(1)_Q$ invariant 
dimension-six operators responsible for the flavour changing 
$\bsll$ transition. The superscript $\ell$ denotes the lepton flavour in the 
final state $(\ell=e,\mu)$. In our notation the short-distance contribution
to the WCs is divided into the SM and the NP ones in the 
following way $\hat{C}^{\ell}_i=C^{SM}_{i}+C^{\ell}_i$. In our analysis we 
consider the subset of operators which are directly responsible for the 
$\bsll$ decay, namely
\bea 
{\mathcal O}_7 &=& \frac{e}{16\pi^2} \,  
m_b \left (\bar s \sigma_{\alpha \beta} P_R b \right )F^{\alpha \beta} \; , 
\\ 
{\mathcal O}^{\ell}_9 &=& \frac{\alpha_{\rm em}}{4 \pi} \,  
\left (\bar s \gamma_\alpha P_L b \right ) (\bar \ell \gamma^\alpha \ell) 
\; , \; \rm \\ 
{\mathcal O}^{\ell}_{10} &=& \frac{\alpha_{\rm em}}{4 \pi} \, 
\left (\bar s \gamma_\alpha P_L b \right ) 
(\bar \ell \gamma^\alpha \gamma_5 \ell) 
\eea
and the set of chirality flipped operators $\mathcal{O}^{'}_i$ obtained 
by interchanging the chiral projectors $(P_L \leftrightarrow P_R) $ in 
the quark current of $\mathcal{O}_i$. The full list of dimension six 
operators also includes scalar, pseudo-scalar and tensor 
structures. However, it has been shown that scalar, pseudo-scalar and tensor 
operators cannot easily give sufficiently large deviations from the SM in the observable 
$R_K$ once constraints from the other $\bsll$ processes are taken into 
account \cite{Alonso:2014csa,Hiller:2014yaa}. Therefore, we will neglect 
these operators in our analysis. Furthermore, we do not consider NP 
contributions to the photonic dipole operator ${\mathcal O}_7$ as it is 
lepton flavour blind by construction. In what follows we assume that all 
the WCs are evaluated at the scale $\mu=m_b$ with the corresponding 
SM contributions given in table \ref{tab:input}.

Apart from $R_K$ we also consider data for other processes which proceed 
via a $\bsll$ transition e.g., the branching ratios of the fully leptonic 
decays $\Bsmumu$ and $\Bsee$, the inclusive decays $\BXsmumu$ and 
$\BXsee$ as well as the branching ratio for the semileptonic decay 
$\BKmumu$. The SM predictions for these branching fractions and their 
current experimental values are summarized in table \ref{tab-1}. 
Note that the experimental upper bound on {$\cal B$}($\Bsee$) is larger than the SM 
prediction by many orders of magnitude. We include this decay mode for completeness but it has no impact on
our final results.
\begin{table}[!h]
\begin{center}
\begin{tabular}{|c|cr|cr|} 
\hline 
Observable & SM prediction &  & Measurement  &  \\
\hline 
$\mathcal{B} \left( B^+ \to K^+ \mu^+ \mu^-\right)_{[1,6]}$ 
& $ \left(1.75^{+0.60}_{-0.29} \right) \times 10^{-7}$   & \cite{Bobeth:2012vn}  
& $ \left(1.21 \pm 0.09 \pm 0.07 \right) \times 10^{-7}$ & \cite{Aaij:2012vr}  \\
$\mathcal{B} \left( B^+ \to X_s \mu^+ \mu^-\right)_{[1,6]}$ 
& $ \left( 1.59 \pm 0.11 \right) \times 10^{-6}$   & \cite{Huber:2005ig}
& $\left( 0.66^{+0.82+0.30}_{-0.72-0.24} \pm 0.07  \right) \times 10^{-6}$ 
& \cite{Lees:2013nxa} \\
$\mathcal{B} \left(B_s \to \mu^+ \mu^-\right)$ 
& $ \left( 3.65 \pm 0.23 \right) \times 10^{-9}$ & \cite{Bobeth:2013uxa}
& $\left( 2.9 \pm 0.7  \right) \times 10^{-9}$ & \cite{CMS-PAS-BPH-13-007} \\
\hline
$\mathcal{B} \left( B^+ \to X_s e^+ e^-\right)_{[1,6]}$ 
& $ \left( 1.64 \pm 0.11 \right) \times 10^{-6}$ & \cite{Huber:2005ig}
& $\left( 1.93^{+0.47+0.21}_{-0.45-0.16} \pm 0.18  \right) \times 10^{-6}$ 
& \cite{Lees:2013nxa} \\
$\mathcal{B} \left( B_s \to e^+ e^-\right)$ 
& $ \left( 8.54 \pm 0.55 \right) \times 10^{-14}$ & \cite{Bobeth:2013uxa}
& $ < 2.8  \times 10^{-7}$ & \cite{Beringer:1900zz} \\
\hline
${R_K}_{[1,6]}$
& $1.0003 \pm 0.0001 $&  \cite{Bobeth:2007dw}
&  $0.745^{+0.090}_{-0.074} \pm 0.036$  & \cite{Aaij:2014ora} \\
\hline
\end{tabular}
\caption{The observables used in our analysis along with their SM 
predictions and experimental measurements.\label{tab-1}}
\end{center}
\end{table} 

We have not used the branching fraction for the decay $\BKee$ 
from \cite{Aaij:2014ora} because of its correlation with the 
$R_K$ measurement. Instead, we have used the value of ${\cal 
B}(\BKmumu)$ from a LHCb measurement in \cite{Aaij:2012vr}. 
Moreover, we have used the data  for ${\cal 
B}(\BKmumu)$ only in the low-$q^2$ bin $1 -6$ GeV$^2$, the main reason being that 
a resonance structure in the dilepton invariant mass distribution was 
observed around $m_{\mu \, \mu}^2 = 17.3$ GeV$^2$ by the LHCb 
collaboration last year \cite{Aaij:2013pta}. This means that even though form factors in the high-$q^2$ region 
were recently computed from lattice QCD \cite{Bouchard:2013mia,Bouchard:2013eph,Horgan:2013hoa,Horgan:2013pva}, the theoretical 
prediction for this observable is affected by non-factorisable hadronic uncertainties.
Information coming from high-$q^2$ measurements of 
this and other observables (such as ${\cal 
B} (B_s \to \phi \mu^+ \mu^-), \, {\cal 
B} (\overline{B}^0 \to \overline{K}^0 \mu^+ \mu^-)$, etc.) can still be used taking into account conservative estimates of the hadronic uncertanties.
We do not expect that the inclusion of such data would change our results drastically, however we expect that the allowed ranges of the WCs would shrink 
due to the extra observables. We also expect a shift of the best fit points in the direction of the SM values, since these extra measurements are generally in good agreement with the SM predictions.



\section{Results}
\label{results}

In this section we present our results for the various NP scenarios 
considered in this paper. As mentioned earlier, we follow a Bayesian 
statistical approach to quantify our results. The details of our 
procedure is explained in the Appendix \ref{procedure}.

\subsection{Single Wilson Coefficient}
\label{results-1}

To start with, we consider only one real NP WC at a time. We will show our 
results for the WCs both in the standard basis (vector and axial-vector 
operators for the lepton current) as well as in the chiral basis for the 
lepton currents. The inclusion of the second set is motivated by the 
possibility of having the NP WCs generated in an $\rm SU(2)_L$ 
invariant way. This possibility was also highlighted in 
\cite{Hiller:2014yaa}.

Our results are summarized in table \ref{tab-2}. The 68\% confidence 
level (C.L.) regions of the WCs are shown in the second column, 
the best fit values are shown in the third column while the last 
column shows the BF for each hypothesis taking the same for the 
$C_9^{\mu}$-only hypothesis as the reference value. More precisely, the 
BF for the hypothesis with NP in the WC $C^{\ell}_i$ is defined as 
\begin{equation} 
{\rm BF}(C^{\ell}_i) = \frac{\int \mathcal{L} \left( 
\textrm{data} | C^{\ell}_i \right) \times P_0(C^{\ell}_i) \, dC^{\ell}_i 
} {\int \mathcal{L} \left( \textrm{data} | C^{\mu}_9 \right) \times 
P_0(C^{\mu}_9) \, dC^{\mu}_9 } \, , 
\end{equation} 
where $\mathcal{L}$ is the likelihood function and $P_0$ is our choice 
of prior for the WC $C^{\ell}_i$, which we assume to be a flat 
distribution in the range [-10,\, 10].

\begin{table}[!h]
\begin{center}
\begin{tabular}{|c|c|c|c|}
\hline
Hypothesis                         & Fit                      & Best fit        & BF                     \\
\hline
$C_9^{\mu}$                        & [-3.1,-0.7]              & -1.6                & $1:1$                    \\
$C_9^{\mu'}$                       & [-1.9,-0.8]              & -1.3                & $0.20:1$                 \\
$C_{10}^{\mu}$                     & [0.7,1.3], [7.5,8.1]     & 1.0                & $0.82:1$                 \\
$C_{10}^{\mu'}$                    & [0.2,0.7]                &  0.5               & $4.8 \times 10^{-3}:1$   \\
$C_9^{\mu} = + C^{\mu}_{10}$       & [0.1,0.8]                &  0.5               & $2.7 \times 10^{-3} : 1$ \\
$C_9^{\mu} = - C^{\mu}_{10}$       & [-0.8,-0.4]              &  -0.6               & $0.42:1$                 \\
$C_9^{\mu'} = + C^{\mu'}_{10}$     & [-0.4,0.3]               &  -0.1               & $9.3\times10^{-4}:1$     \\
$C_{9}^{\mu'} = - C_{10}^{\mu'}$   & [-0.2,-0.6]              &  -0.4               & $1.3\times10^{-2}:1$     \\
\hline
$C_9^{e}$                          & [-8.4,-8.4], [0.6,2.1]   &   1.3              & $0.13 : 1$               \\
$C_9^{e'}$                         & [0.8,1.9]                &   1.3              & $0.10:1$                 \\
$C_{10}^{e}$                       & [-1.6,-0.7], [9.5,10.0]  &   -1.1              & $0.14 : 1$               \\
$C_{10}^{e'}$                      & [-1.7,-0.7]              &   -1.1              & $9.7 \times10^{-2}:1$    \\
$C_9^{e} = + C^{e}_{10}$           & [-2.4,-1.4], [2.2,3.4]   &   -1.9              & $0.20 : 1$               \\
$C_9^{e} = - C^{e}_{10}$           & [0.3,1.1]                &  0.6               & $6.7 \times10^{-2} : 1$  \\
$C_9^{e'} = + C^{e'}_{10}$         & [-2.6,-1.5], [2.2,3.2]   &  -2.0               & $0.20:1$                 \\
$C_{9}^{e'} = - C_{10}^{e'}$       & [0.4,0.9]                &   0.7              & $5.2\times10^{-2}:1$     \\
\hline
$C_9^{\mu}$=$C_9^{e}$              & [-4.3,-1.1]              &      -2.2           & $2.9\times10^{-2}:1$                  \\
$C_{10}^{\mu}$=$C_{10}^{e}$        & [0.3,1.2], [7.6,8.4]     &      0.8           & $1.7\times10^{-2}:1$                  \\
\hline
SM                                 &                          &                 & $2.4 \times 10^{-3}:1$   \\ 
\hline
\end{tabular}
\caption{The 68\% C.L. ranges for the WCs when only one WC is considered 
at a time. The data on $\BKstarmumu$ have not been used at this stage.}
\label{tab-2}
\end{center}
\end{table}

\begin{figure}[t!]
\begin{center}
\begin{tabular}{cc}
\includegraphics[scale=0.6]{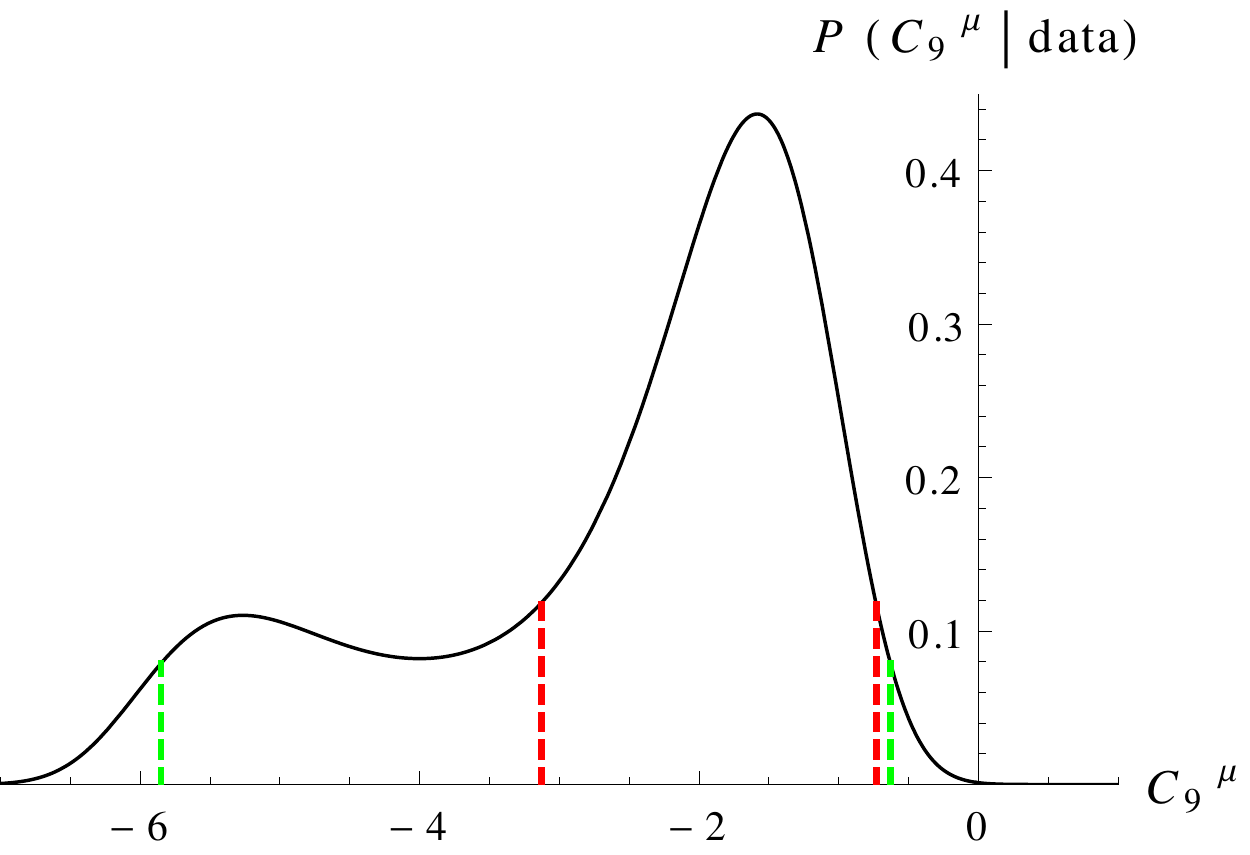} & 
\includegraphics[scale=0.6]{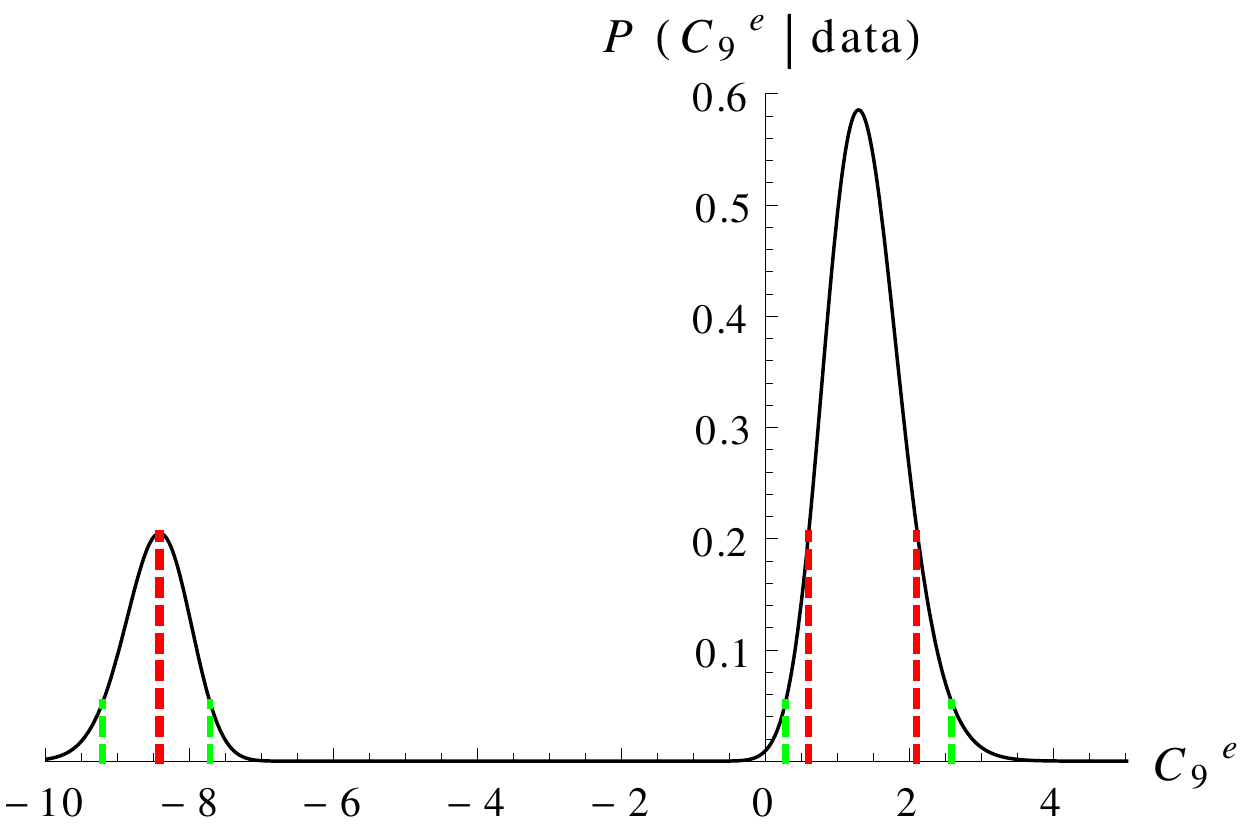}
\end{tabular}
\caption{Posterior probabilities for the $C_9^{\mu}$ only (left) and $C_9^{e}$ 
only (right) hypotheses. The red and green vertical lines refer to boundaries 
of 68\% and 95\% regions respectively.
\label{fig-1}}
\end{center}
\end{figure}

In Figure \ref{fig-1} we show the posterior probabilities for the 
$C_9^{\mu}$ only and $C_9^{e}$ only hypotheses as examples.

We now discuss some general features of our results shown in table 
\ref{tab-2}. Clearly, the hypothesis with non-zero $C^{\mu}_9$ is the most favoured by the data as it offers the largest BF. The $C^{\mu}_{10}$ NP scenario 
also does quite well. As $C^{\mu \, \prime}_9$ does not contribute to 
the decay $\Bsmumu$ (whose experimental central value is now slightly 
lower than the SM prediction), its BF is reduced to some extent. The 
extremely low BF for the $C^{\mu'}_{10}$ case is due to a tension 
between $R_K$ and and $\mathcal{B} \left(\Bsmumu\right)$. It can be seen 
from the expressions of these two observables (see Appendices \ref{bkll} 
and \ref{bsll} ) that the experimental value for $R_K$ prefers 
$C^{\mu'}_{10}>0$ while the measured branching ratio for $\Bsmumu$ 
prefers the opposite. The hypotheses $C^{\mu (')}_{9} = C^{\mu (')}_{10} 
$ (which correspond to the operator directions $(\bar s \gamma_\alpha 
P_L b ) (\bar \ell \gamma^\alpha P_R\ell)$ and $(\bar s \gamma_\alpha 
P_R b ) (\bar \ell \gamma^\alpha P_R\ell)$) are also strongly 
disfavoured because they generate $R_K \gtrsim 1$ which is in tension with 
experiment. The other chiral operator $(\bar s \gamma_\alpha P_L b ) 
(\bar \ell \gamma^\alpha P_L\ell)$ (our hypothesis 
$C^{\mu}_9=-C^{\mu}_{10}$) turns out to be the best among the four 
chiral operators. This case was also considered in  
Ref.~\cite{Hiller:2014yaa} in the context of their model independent analysis 
as well as a specific leptoquark model. In their analysis they quote
$C^{\mu}_9=-C^{\mu}_{10} \approx 
-0.5$ as a benchmark point which is in fact consistent with our 68\% CL range in table 
\ref{tab-2}.

As far as NP in the electron sector is concerned, all the hypotheses 
give similar BFs (at most a factor 4 between the best and the worst). 
This can be understood by the fact that the measurements with electrons in final states have 
a larger experimental error than those with muons in the final state.
Note also that there are a few cases with multiple solutions (because the 
rates are quadratically dependent on the WCs), some of them with very 
large NP contributions (see for example $C^e_{10}$). Including other 
observables (for example those in the decay $B \to K^* \ell^+ \ell^-$) 
will certainly modify this picture.

The scenario $C^{e '}_{10}=-C^{e '}_{9} \approx 0.5$ was also considered 
in the Ref.~\cite{Hiller:2014yaa}. Although their estimate $C^{e 
'}_{10} \approx 0.5$ is compatible with the 68\% CL region from our fit, 
its BF is the worst among the hypotheses with NP in the electron sector.

The highest BFs are obtained for hypotheses with NP in the muon sector.
In our fit, this finding is driven by the observables $\mathcal{B}(B \rightarrow X_s \ell^+ \ell^-)$ and $\mathcal{B}(B^+ \rightarrow K^+ \mu^+ \mu^-)$, for which the dimuon channel measurements are 
each lower (at more than 1$\sigma$ level) than the SM predictions, while the electron channel measurement is in good agreement
 with the SM. 
Hence our fit finds a slight
 preference for NP in the muon sector. 
For example, a negative value for $C_9^{\mu}$ will lower the predictions for both $\mathcal{B}(B \rightarrow X_s \mu^+ \mu^-)$ and $\mathcal{B}(B^+ \rightarrow K^+ \mu^+ \mu^-)$ 
compared to the SM while keeping $\mathcal{B}(B \rightarrow X_s e^+ e^-)$ at its SM value.

Finally, we show our results for the two lepton universal cases 
$C_9^{\mu}$=$C_9^{e}$ and $C_{10}^{\mu}$=$C_{10}^{e}$ (see the last but 
one row in table \ref{tab-2}). It can be seen that the BF for both 
these two cases are rather low compared to most of the non-universal NP 
scenarios in particular, the $C_9^{\mu}$-only and $C_{10}^{\mu}$-only 
hypotheses.


\subsection{Combination of two Wilson Coefficients at a time}
\label{results-2}

In this section we allow the possibility of having NP in two WCs 
simultaneously and study the consequences. Here we do not 
consider all the possible combinations of the WCs, rather, we take the 
few best cases from table \ref{tab-2} and consider their combinations. 
Our results are summarized in table \ref{tab-3}. The 68\% range of a 
parameter is obtained after marginalizing over the other parameter.

The results of our analysis show that, in general, there is no particular 
gain in considering NP effects  in two WCs at the same time, 
this is more marked in the case of the chiral operators.

\begin{figure}[t!]
\begin{center}
\begin{tabular}{cc}
\includegraphics[width=7cm]{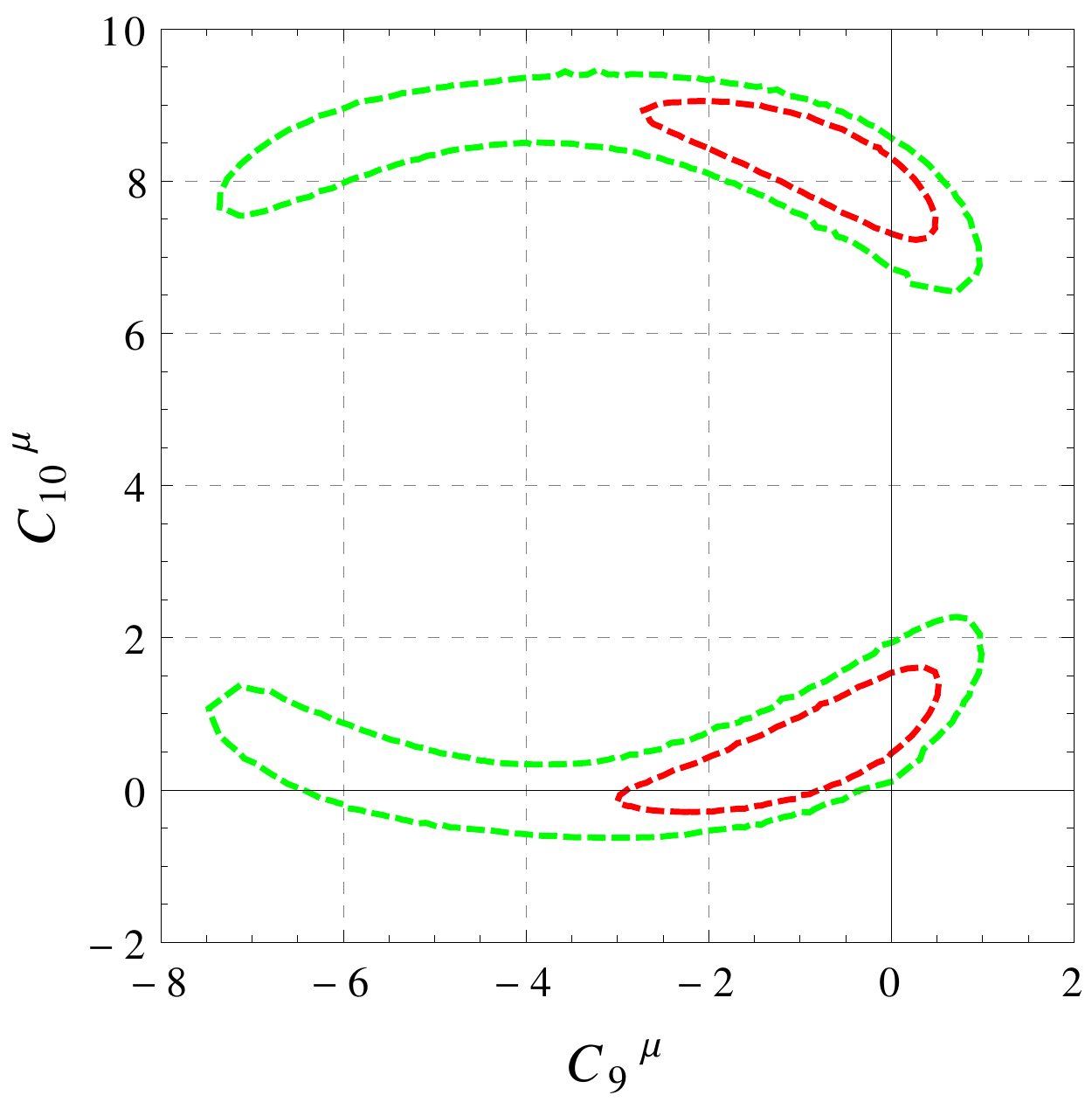} &
\includegraphics[width=7cm]{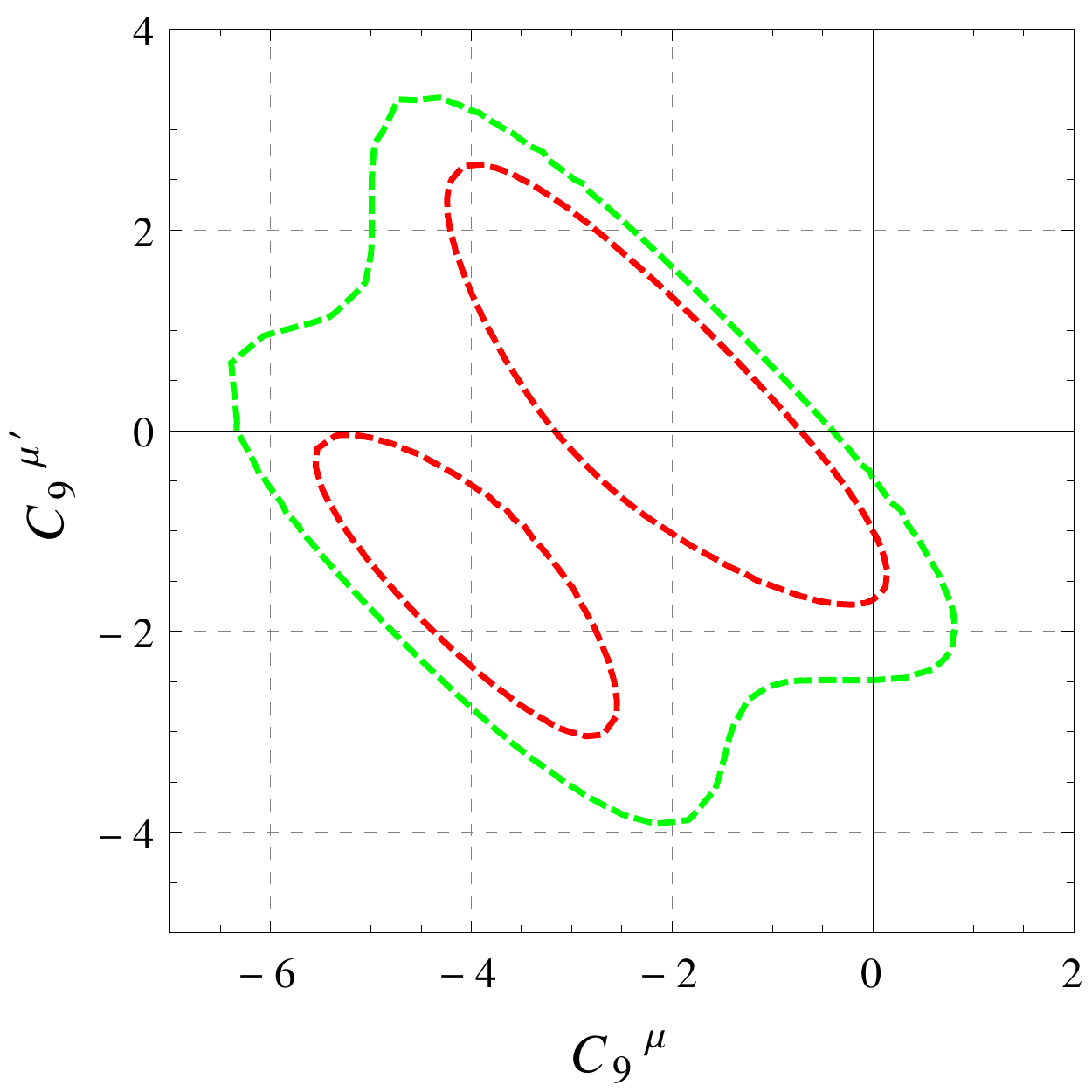}
\end{tabular}
\caption{Two dimensional posterior probability distributions for the two hypotheses:
1. $C_9^{\mu}$ and $C_{10}^{\mu}$ (left panel), 
2. $C_9^{\mu}$ and $C_{9}^{\mu \, \prime}$ (right panel). 
The red and green contours are 68\% and 95\% C.L. regions respectively. 
\label{fig-2a}}
\end{center}
\end{figure}

Among the hypothesis of NP in two WCs, the largest BF is obtained for the pair $ \{ C_9^e,C^{\mu}_9 \} $.
The 2-dimensional posterior distribution for this scenario 
is shown in the left panel of figure \ref{fig-2a}. Similarly, the 
posterior probability distribution for the NP scenario with both 
$C_9^{\mu}$ and $C_{9}^{\mu \, \prime}$ is shown is the right panel of 
figure \ref{fig-2a}. It can be seen that the data is consistent with no 
NP in $C_{9}^{\mu \, \prime}$. 

\begin{table}[t!]
\begin{center}
\begin{tabular}{|c|c|c|c|}
\hline
Hypothesis                   & Fit                       & Best fit        & BF          \\
\hline
$C_9^{\mu}$                  & [-1.9,0.3]                & -0.6                &     $0.15:1$        \\ 
$C_{10}^{\mu}$               & [-0.1,0.9], [8.0,8.8]     & 0.4                &            \\
\hline
$C_9^{\mu}$                  & [-4.2,-1.2]               & -2.8                &     $0.20:1$        \\
$C_{9}^{\mu'}$               & [-1.7,1.2]                & -0.3                &            \\
\hline
$C_9^{\mu}$                  & [-4.2,-1.4]               &  -2.6               &      $0.28:1$       \\
$C_9^{e}$                    & [-7.4,-5.9], [-1.3,0.2]   &  -6.6               &             \\
\hline
$C_9^{\mu}=-C_{10}^{\mu}$    & [-1.0,0.4]                &  -0.7               &       $4.5 \times 10^{-2}:1$      \\ 
$C_9^{e} = -C_{10}^{e}$      & [-0.5,0.4], [-8.2,-7.4]   &  -0.1               &             \\
\hline
$C_9^{\mu}=-C_{10}^{\mu}$    & [-0.7,-0.4]               &  -0.5               &       $8.3 \times 10^{-2}:1$      \\
$C_9^{e} = C_{10}^{e}$       & [-1.2,1.6]                &   -0.2              &             \\
\hline
$C_9^{\mu}=C_{10}^{\mu}$     & [0.1,0.9]                 &   0.5              &      $8.0 \times 10^{-3}:1$       \\ 
$C_9^{e} = -C_{10}^{e}$      & [0.3,1.1]                 &   0.6              &             \\
\hline
$C_9^{\mu}=C_{10}^{\mu}$     & [0.1,0.9]                 &   0.5              &     $2.4 \times 10^{-2}:1$        \\
$C_9^{e} = C_{10}^{e}$       & [-2.4,-1.5], [2.2,3.4]    &   2.8              &             \\
\hline
\end{tabular}
\caption{Same as table \ref{tab-2} but with two WCs turned on simultaneously.}
\label{tab-3}
\end{center}
\end{table}

\begin{figure}[t!]
\begin{tabular}{cc}
\includegraphics[width=7cm]{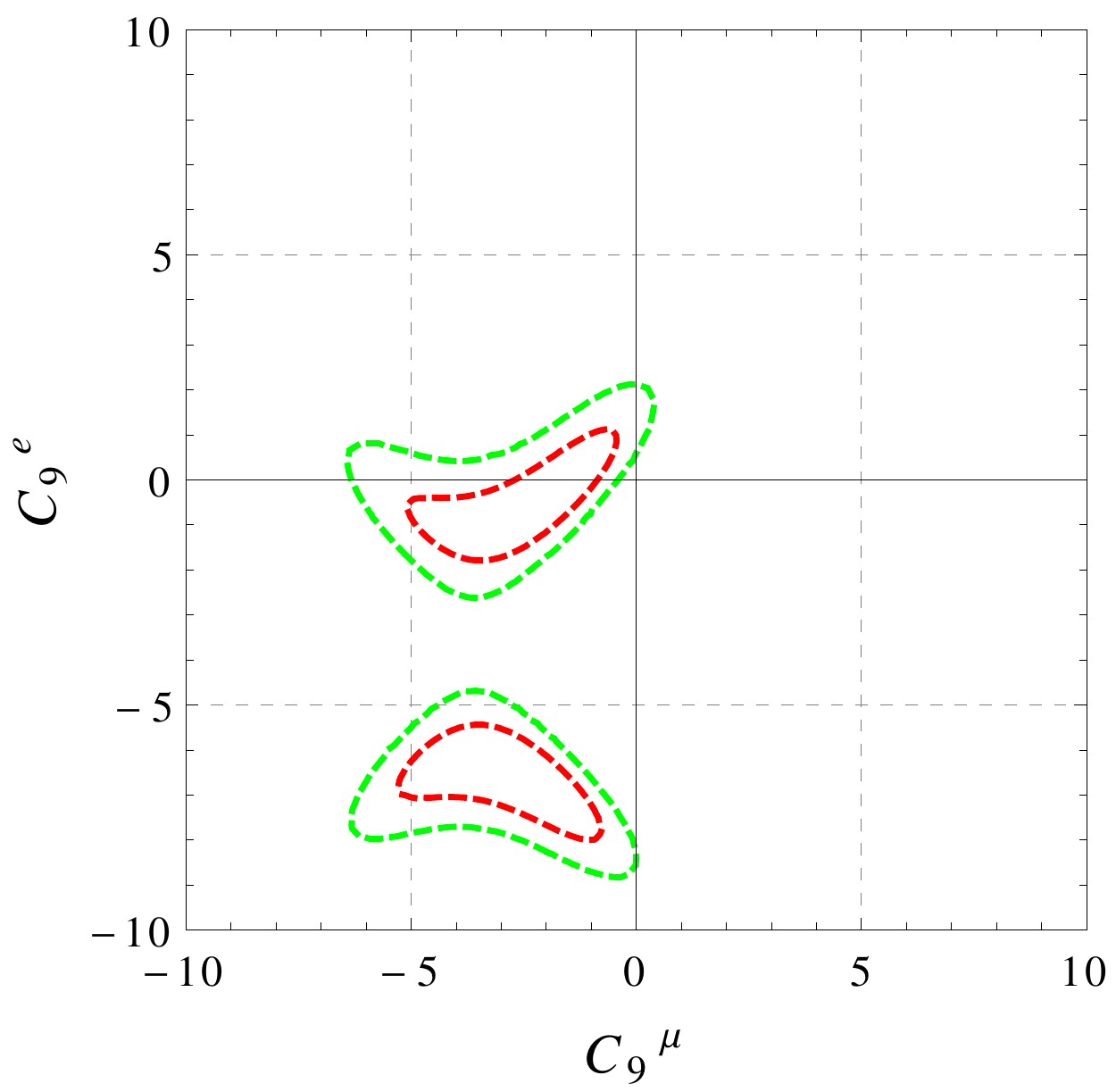} & 
\includegraphics[width=7cm]{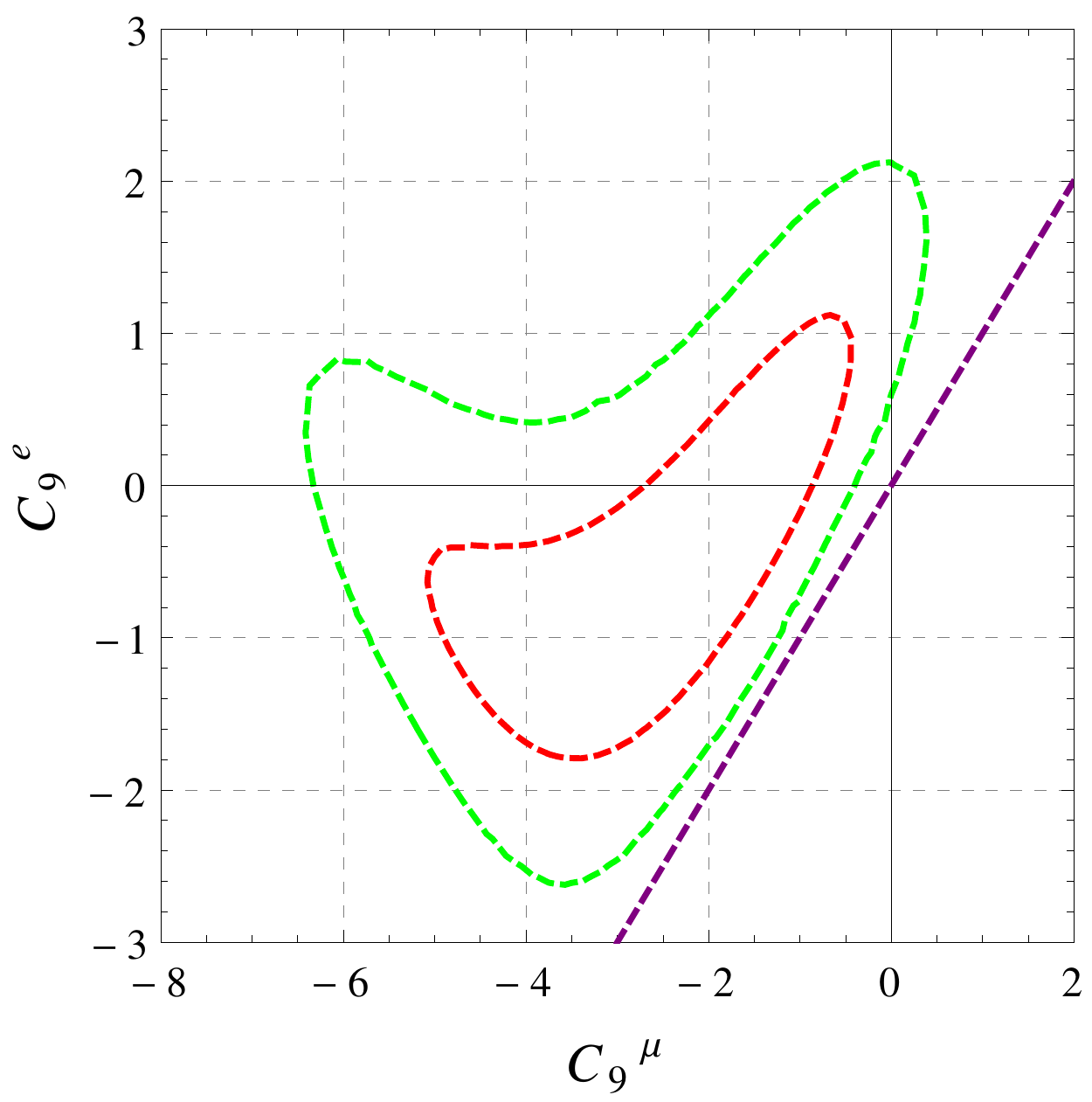}
\end{tabular}
\caption{Two dimensional posterior probability distribution for the 
hypothesis with NP in $C_9^{\mu}$ and $C_{9}^{e}$. In the left panel 
both the allowed regions are shown. In the right panel we show only the 
region close to the origin. The red and green contours are 68\% and 95\% 
C.L. regions respectively. The purple line represents the lepton flavour 
universal scenario $C_9^{\mu} = C_{9}^{e}$. \label{fig-2b}}
\end{figure}

The posterior probability for the hypothesis with both $C_9^{\mu}$ and 
$C_{9}^{e}$ turned on is shown in figure \ref{fig-2b}. In the left panel 
we show both the allowed regions, while in the right panel we show a 
zoomed in version of the region close to the SM point $C_9^{\mu} = 
C_{9}^{e} =0$. Figure \ref{fig-2b} clearly shows that the data prefers 
NP in the muon sector over NP in the electron sector. Moreover, NP with 
lepton flavour universality (shown by the the dashed purple line) is 
disfavoured by more than 95\% C.L. 

\subsection{Including the data on $B \to K^* \mu^+ \mu^-$}
\label{BKsll}

The latest LHCb measurement of the decay distribution in $\BKstarmumu$ 
has seen interesting deviations in several observables from their SM 
predictions \cite{Aaij:2013iag,Aaij:2013qta}. These deviations were most 
pronounced in two of the so-called optimized observables (where the 
hadronic uncertainties are expected to cancel to a good extent) 
$P_5^{\prime}$ and $P_2^{\prime}$ in the low-$q^2$ region~\cite{Descotes-Genon:2013vna}. 
Soon after 
the LHCb result was published, it was shown in 
\cite{Descotes-Genon:2013wba} that a good fit to all the data can be 
obtained by having a negative contribution in the range $[-1.9, -1.3]$ to the 
WC ${\hat C}_9^\mu$. A similar conclusion was also reached by two other groups 
 \cite{Altmannshofer:2013foa,Beaujean:2013soa}. The fit done in 
\cite{Altmannshofer:2013foa} found a need for a NP contribution 
(similar to $C_9^\mu$ in magnitude but with opposite sign) also to the 
chirally flipped operator ${\hat C}_9^{\mu \, \prime}$. As discussed in section 
\ref{results-2} this is now in tension with the measurement of $R_K$ 
(assuming the presence of NP only in the muon sector).

Note that the issue of hadronic uncertainties, in particular the role of 
long-distance $c \bar{c}$ loops still remains unclear, see 
\cite{Descotes-Genon:2014uoa} and the references therein for a recent 
discussion. Thus, the jury is still out on whether NP has already been 
seen in these measurements. Despite this uncomfortable situation, 
several NP interpretations of the LHCb data have been proposed 
\cite{Gauld:2013qja,Datta:2013kja,Buras:2013qja} and it would be 
interesting to see whether the $R_K$ measurement can be reconciled with 
the $\BKstarmumu$ data. With this motivation, in this section we combine 
the result of \cite{Descotes-Genon:2013wba} with our analysis.

\begin{table}[t!]
\begin{center}
\begin{tabular}{|c|c|c|c|}
\hline
Hypothesis           & Fit                          & Best Fit          & BF       \\
\hline
$C_9^{\mu}$          & [-1.9,-1.3]                  &   -1.6                & $1:1$    \\ 
\hline
$C_9^{\mu}$          & [-1.9,-1.3]                  &   -1.6                & $0.14:1$ \\
$C_{9}^{e}$          & [-7.7,-6.6], [-0.7,0.6]      &   -0.1                &          \\
\hline
$C_9^{\mu}$          & [-1.8,-1.4]                  &   -1.6                & $0.13:1$ \\
$C^e_{10}$           & [-0.4,0.5], [8.3,9.3]        &   8.7                &          \\
\hline
$C_9^{\mu}$          & [-1.8,-1.3]                  &  -1.5                 & $0.16:1$ \\
$C_{9}^{e}=C^e_{10}$ & [-0.9,1.5]                   &  -0.1                 &          \\
\hline
$C_9^{\mu}$           & [-1.9,-1.3]                 &  -1.6                 & $6.0 \times 10^{-2}:1$  \\
$C_{9}^{e}=-C^e_{10}$ & [-8.2,-7.8], [-0.3,0.3]     &  0.0                 &          \\
\hline
\end{tabular}
\caption{Results when the data on $\BKstarmumu$ are also included 
as discussed in section \ref{BKsll}. The same conventions as in 
table \ref{tab-2} and \ref{tab-3} are used.}
\label{tab-4}
\end{center}
\end{table}

We follow an approximate procedure (see Appendix \ref{procedure} for 
more details) which allows us to use their result where NP only in 
${\hat C}^{\mu}_9$ and ${\hat C}_{7}$ was considered, see their 
eq. 4. Note that, although the analysis in \cite{Descotes-Genon:2013wba} 
was performed assuming lepton universality their results can, to a very 
good approximation, be taken as valid only for the muonic WCs. This 
allows us to safely use their result with NP in $\hat{C}^{\mu}_{9}$ and 
any operator in the electron sector.  As we do not consider NP in the 
dipole operator in this paper, $\hat{C}_{7}$ is set to its SM 
value (which is consistent with the 68\% C.L. region of 
\cite{Descotes-Genon:2013wba}).

\begin{figure}[h!]
\begin{tabular}{cc}
\includegraphics[width=7cm]{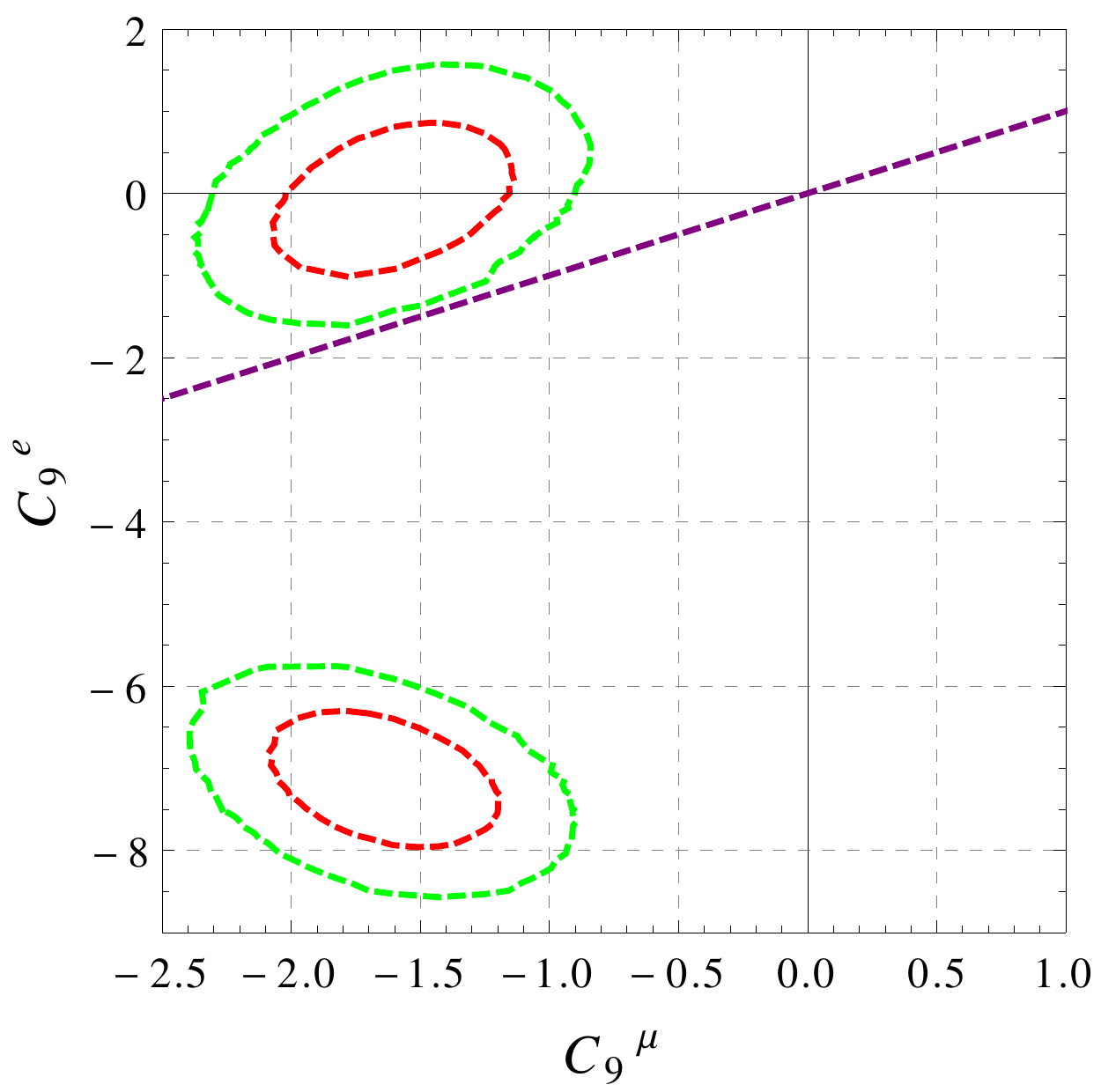} & 
\includegraphics[width=7cm]{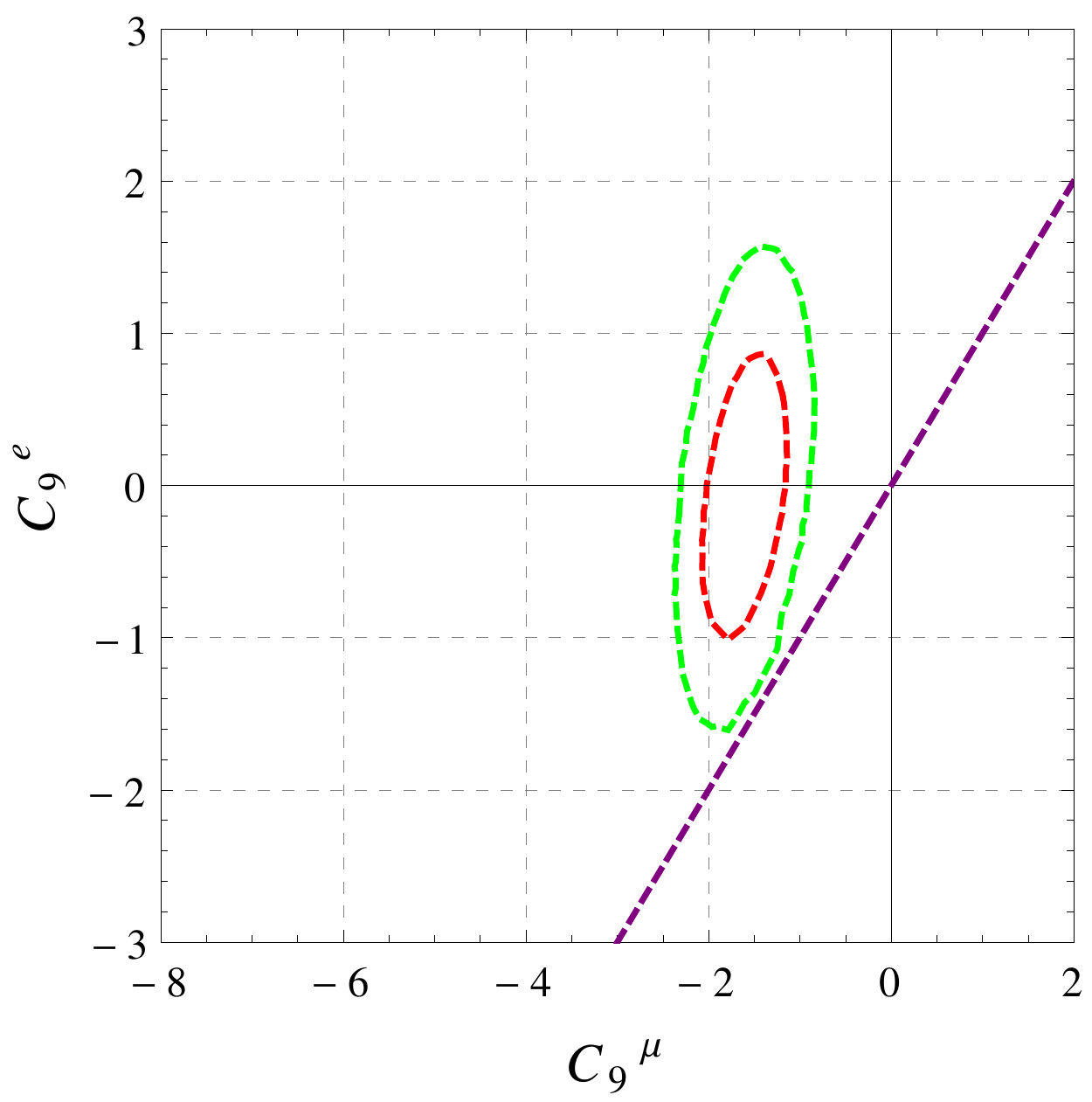}
\end{tabular}
\caption{Two dimensional posterior probability distributions for the hypothesis 
$C_9^{\mu}$ -- $C_{9}^{e}$ including the data on $\BKstarmumu$ 
(see text for more details). The same conventions as in figure \ref{fig-2b} are used. 
\label{fig-3}}
\end{figure}

The result of our fit is reported in table \ref{tab-4}. In the first row 
we show the result when only $C^{\mu}_{9}$ is turned on while in the 
following rows we allow NP in the electron sector in addition to 
$C^{\mu}_{9}$. We notice that the range for $C^{\mu}_{9}$ preferred by 
the analysis of \cite{Descotes-Genon:2013wba} is confirmed even with the 
inclusion of our observables. Allowing the possibility of NP in the 
electron sector makes things slightly better (increases the BF by 
roughly a factor of 3), except for the scenario of the last row of the table 
where the BF remains almost the same. The posterior probability 
distribution for the $C^{\mu}_{9}$ - $C^{e}_{9}$ hypothesis is shown in 
figure \ref{fig-3}. Similar to figure \ref{fig-2b}, in the left panel 
both the allowed regions are shown while in the right panel only the 
region close to the SM point is shown. Again, the preference for 
lepton flavour non-universal NP is very clear. The data is consistent 
with no NP in the electron sector but demands a rather large NP in the 
muon sector. A comparison of the figure in the right panel with that in 
figure \ref{fig-2b} will also show the constraining power of the 
$\BKstarmumu$ data. We warn the readers that the BFs in the table 
\ref{tab-4} should not be compared with those in the previous section 
but should only be compared within table \ref{tab-4}. This is because 
the analysis in this section uses different set of information than 
those in the previous section. Also, we are combining our analysis 
with that in \cite{Descotes-Genon:2013wba} in a very simplified way.
We report the BFs just to show that the data 
does not show a strong preference for any particular kind of NP in the 
electron sector; many of them do equally well.

\section{Summary and conclusions}
\label{conclusion}

The flavour changing process $\bsll$ is responsible for many rare decays 
such as $\Bsll$, $\BXsll$, $\BKll$ and $\BKstarll$. These decays, being 
extremely rare in the SM, are very powerful probes of NP. 
The LHCb 
collaboration already made significant progress last year by measuring most of 
the observables in the full three angle distribution of the decay 
$\BKstarmumu$. Interesting deviations from the SM predictions were 
seen in a few of the so-called ``form-factor independent'' or optimized 
observables. Although several NP explanations of these deviations were 
put forward, firm confirmation of NP in these observables is not yet 
possible due to the hadronic uncertainties which are not completely understood.
The $2.6 \sigma$ deviation in $R_K$, although not yet statistically 
significant, is worthy of attention because the ratio $R_K$ is 
essentially free of hadronic uncertainties in the SM. Moreover, any 
lepton flavour blind new short distance physics would predict $R_K 
\approx R_K^{\rm SM}$ and hence, the confirmation of this deviation 
would clearly point towards lepton flavour non-universal NP. This has 
motivated us to study model independent NP explanations of the
measurement of $R_K$ considering various  
observables involving a $\bsll$ transition. To this end, we have 
performed a Bayesian statistical analysis of the various NP scenarios. 
We have also quantified the goodness-of-fit of these NP hypotheses by 
computing and comparing their Bayes Factors.
 
We first performed a fit without including the data on 
$\BKstarmumu$. Our results for the hypothesis of NP in one (two) WC(s) at a 
time are summarized in Table \ref{tab-2} (\ref{tab-3}). 
In the muon sector, among the fits to a single 
WC in the standard basis, 
only $C_9^{\mu}$ and $C_{10}^{\mu}$ have large BFs while among the chiral 
operators, the hypothesis of NP in $C_9^{\mu} = - 
C_{10}^{\mu}$ gives the highest BF. 
In the electron sector 
all the NP hypotheses give comparable BFs, although worse than 
the scenarios with $C_9^{\mu}$ or $C_{10}^{\mu}$. Even without the inclusion of the $\BKstarmumu$ data, our fit
shows a slight preference towards the hypothesis of NP in the di-muon sector. 
However it should be noted that large NP 
effects in the electron sector are not excluded and in the future, with more precise measurements, 
the situation could change. We further show
that the lepton flavour universal NP scenarios for example, $C_9^{\mu} = 
C_{9}^{e}$ or $C_{10}^{\mu} = C_{10}^{e}$ have rather low BFs and hence they are disfavoured.

When two NP WCs are turned on simultaneously, the situation does not particularly improve. 
We have shown a few posterior distributions in 
figures \ref{fig-2a} and \ref{fig-2b} and we notice that for the 
$C_9^{\mu}$--$C_9^{e}$ hypothesis the lepton flavour universal case
$C_9^{\mu} = C_9^{e}$ is disfavoured at more than 95\% C.L. We continued in section \ref{BKsll} by including also the data on 
$\BKstarmumu$. We employed a simplified procedure (explained in appendix 
\ref{procedure}) to combine part of the results from 
\cite{Descotes-Genon:2013wba} with our observables. We observed that the 
allowed range of $C^{\mu}_{9}$ obtained in \cite{Descotes-Genon:2013wba} 
remains consistent with the $R_K$ data. Our analysis also showed that 
the data is consistent with no NP in the electron sector. 


A global analysis including the final states with tau 
leptons would be very interesting once more data are available. In fact, the 
possibility of large enhancements in many of the $b \to s \tau^+ \tau^-$ 
modes were already discussed in the context of like-sign dimuon 
asymmetry seen at the Tevatron, see for example \cite{Dighe:2010nj,Dighe:2012df}.

\bigskip
{\bf Acknowledgments:}

DG is supported by the European Research Council under the European 
Union's Seventh Framework Programme (FP/2007-2013) / ERC Grant Agreement 
n.279972. DG would like to thank Prof. Benjamin Allanach for his 
hospitality in DAMTP, University of Cambridge where this project was 
envisaged. DG also thanks Prof. Michael Spannowsky and the CERN theory 
division for their hospitalities in IPPP Durham and CERN respectively 
where part of this work was carried out. 
Discussion with Prof. Luca Silvestrini is also gratefully acknowledged. 
MN and SR are supported by STFC. MN and SR want to thank Ben Allanach
for useful discussions on Bayesian statistics. Thanks to Diego Guadagnoli 
and David Straub for helpful comments.


\clearpage
\appendix


\section{Details of the analysis}

\subsection{$B^+ \to K^+ \ell^+ \ell^-$}
\label{bkll}

The branching ratio for the decay $B^+ \to K^+ \ell^+ \ell^-$ in the 
low-$q^2$ region can be written as 
\cite{Bobeth:2007dw,Alok:2010zd,Alok:2011gv,Khodjamirian:2010vf},
\begin{equation}
 \begin{array}{rl}
 \mathcal{B} \left( B^+ \to K^+ \ell^+ \ell^-\right)_{[1,6]} 
 =& \dfrac{\tau_B^{\pm} G_F^2 \alpha^2_e \left| V_{tb} V^*_{ts} \right|^2}
 {\, 2^9 \pi^5 m^3_{B^{\pm}}} \dfrac{1}{3} \times \Big{(} \\
&  \left| C^{\rm SM}_{10} + C^{\ell}_{10} +C^{\ell'}_{10}  
\right|^2  (f^+_{BK})^2 \left( I_1 +2 I_2 \, b^+_1 +I_3 (b^+_1)^2 \right) +  \\
&  \left| C^{\rm SM}_9 +C^{\ell}_{9} +C^{\ell'}_{9}  
\right|^2  (f^+_{BK})^2 \left( I_1 +2 I_2 \, b^+_1 +I_3 (b^+_1)^2 \right) +  \\
&  \left| C^{\rm SM}_7 \right|^2  (f^T_{BK})^2 \left( I_1 +2 I_2 \, b^T_1 +I_3 (b^T_1)^2 \right) 
\left(  \dfrac{2 m_b}{m_{B^{\pm}}+m_{K^{\pm}}} \right)^2 +  \\
& {\rm Re} \left[ C^{\rm SM}_7 (C^{\rm SM}_9 +C^{\ell}_{9} 
+C^{\ell'}_{9})^* \right] \times \\
& \left.  f^+_{BK} f^T_{BK} \left( I_1 + I_2 \,( b^+_1+ b^T_1) +I_3 b^+_1 b^T_1 \right) 
\dfrac{4 m_b}{m_{B^{\pm}}+m_{K^{\pm}}}   \right) \, \, ,
\end{array}
\end{equation}
where the numerical values for $I_1$, $I_2$ and $I_3$ are given by, 
\begin{eqnarray*}
I_1 &=& 89729.8 \, , \\
I_2 &=& -2703.44 \, , \\
I_3 &=& 97.3014 \, . 
\end{eqnarray*}

In order to obtain these numbers we have used the form factor 
parametrization given in \cite{Khodjamirian:2010vf}. The values of the 
form factor parameters $f^+_{BK}$, $f^T_{BK}$, $b_1^+$ and $b_1^T$ are 
given in table \ref{tab:input}. As far as the uncertainties in these 
parameters are concerned, for $b_1^+$ and $b_1^T$ we have taken 
asymmetric gaussian priors in the fit. For the other two parameters 
$f^+_{BK}$ and $f^T_{BK}$, we have fixed them to their respective 
central values and the associated uncertainties are taken into account 
by rescaling the experimental error
$\sigma_{ \mathcal{B}}$ in $\mathcal{B} \left( B^+ \to K^+ \mu^+ \mu^-\right)_{[1,6]}$ by  
\begin{eqnarray}
\sigma_{ \mathcal{B}} & \to &  
\sqrt{\sigma_{ \mathcal{B}}^2 + 4 \frac{\sigma^2_f}{f^2_{BK}} \overline{\mathcal{B}}^2  } 
= 0.37 \times 10^{-7} \,
\end{eqnarray}
(taking conservatively the values $f_{BK}=0.34$, $\sigma_f = 0.05$ and 
$\overline{\mathcal{B}}=1.21 \times 10^{-7}$) . We made this 
(conservative) approximation in order to reduce the computational time 
in the evaluation of the posterior distributions for the WCs and avoid 
some numerical instabilities. We have followed the same prescription 
also for the branching ratio of $\BKee$. The theoretical uncertainties 
coming from $f^+_{BK}$ and $f^T_{BK}$ in the two observables 
${\cal B}(\BKmumu)$ and ${\cal B}(\BKee)$ have been assumed to be 
independent for simplicity.

\subsection{$R_K$}
\label{rk}

The expression for $R_K$ can be derived using the result given in the 
previous subsection. Here we give a simplified formula which can be 
useful for analytic understanding of our results. Observing the fact 
that $\left| \dfrac{C^{SM}_{7}}{C^{SM}_{9,10}} \right| \approx 0.07$ and 
also that the NP in $\hat{C}_7$ is severely constrained by the measured 
branching ratio of $B \to X_s \gamma$, the expression of $R_K$ can be 
approximated by,

\begin{equation}
R_K \approx \dfrac{\left| C^{SM}_{10} + C^{\mu}_{10} +C^{\mu'}_{10} 
\right|^2 +  \left| C^{SM}_9 +C^{\mu}_{9} +C^{\mu'}_{9}  \right|^2}
{\left| C^{SM}_{10} + C^{e}_{10} +C^{e'}_{10}  \right|^2 +  \left| 
C^{SM}_9 +C^{e}_{9} +C^{e'}_{9}  \right|^2} \, . 
\end{equation}

This simplified expression clearly shows that the theoretical error in 
$R_K$ is very small even in the presence of NP. Note that in our numerical 
fit we have used the full expression of $R_K$ with all the form 
factor dependent terms.

\subsection{$B \to X_s \, \ell^+ \, \ell^-$} \label{bxsll} 
%
The branching ratio for the inclusive decay $B \to X_s \, \ell^+ \, 
\ell^-$ in the low-$q^2$ region can be written as 
\cite{DescotesGenon:2011yn}
\begin{eqnarray}
{\cal B}(B \to X_s \ell^+ \ell^-)_{[1,6]} = 10^{-7} \times
\bigg[ (15.86 \pm 1.51)  &+& 2.663 \, C_9^\ell - 0.049 \, C_9^{\ell '}
- 4.679 \,  C_{10}^{\ell} + 0.061 \,  C_{10}^{\ell '} \nonumber \\
&+& 0.534 \, ({ C_9^\ell}^2 + { C_9^{\ell '}}^2)
+ 0.543 \, ({ C_{10}^\ell}^2 + { C_{10}^{\ell '}}^2) \nn \\
&-& 0.014 \, \,  C_9^\ell  C_9^{\ell '} -  0.014 \, \, 
 C_{10}^{\ell}C_{10}^{\ell '}\bigg] \, . 
\end{eqnarray}
We took the theoretical error into account by taking a 
gaussian prior for a parameter called $b_0$ and marginalizing over it, with its 
central value and standard deviation given by 15.86 and 1.51 respectively.

\subsection{$B_s \to \ell^+ \, \ell^-$}
\label{bsll}
The branching ratio for the leptonic decays $\Bsmumu$ and $\Bsee$ can be 
written as
\begin{eqnarray}
{\cal B}(B_s \to \mu^+ \mu^-) &=&
{\cal B}(B_s \to \mu^+ \mu^-)_{\rm SM}
\bigg|\dfrac{C_{10}^{\rm SM}+C_{10}^\mu - C_{10}^{\mu \, \prime}}
{C_{10}^{SM}}\bigg|^2 \, \, \,\\
{\cal B}(B_s \to e^+ e^-) &=&
2.34 \times 10^{-5} \times {\cal B}(B_s \to \mu^+ \mu^-)_{\rm SM}
\bigg|\dfrac{C_{10}^{\rm SM}+C_{10}^e - C_{10}^{e \, \prime}}{C_{10}^{\rm SM}}\bigg|^2 \, \,. 
\end{eqnarray}
Here we have used the SM value of ${\cal B}(\Bsmumu)$ as our input parameter instead 
of the $B_s$ meson decay constant $f_{B_s}$ and the CKM matrix elements. 
We took the theoretical error on this into account by taking a gaussian prior for 
${\cal B}(B_s \to \mu^+ \mu^-)_{\rm SM}$ and marginalizing over it, using values 
from \cite{Bobeth:2013uxa}:
$${\cal B}(B_s \to \mu^+ \mu^-)_{\rm SM} = (3.65 \pm 0.23) \times 10^{-9} \, .$$

\section{Statistical procedure}
\label{procedure}

In this appendix we briefly describe the statistical procedure that has 
been followed in this work. Our aim is to construct the posterior 
probability distributions (p.d.f.) for a set of WCs that we denote 
collectively by $\mathbf{C}$, for example $\mathbf{C} = \{ C_9^{\mu}, 
C_9^e, .... \}$.

The theoretical prediction for a given observable will be denoted by $O$ 
and it will depend on $\mathbf{C}$ as well as a set of input parameters 
$\mathbf{x}$, for example CKM matrix elements, form factors, masses of 
the particles etc. The inputs are either experimentally measured 
quantities or theoretically calculated parameters with some 
uncertainties. In order to take into account these uncertainties we have 
to attach a probability distribution $f_{x_i}(x_i)$ to each input $x_i$.

The measured observables will be assumed to be distributed according to a 
Normal distribution with mean value $\overline{O}$ and standard 
deviation $\sigma_{O}$. The generalization to distributions other than the
Normal distribution is straightforward.

If one has a set of $N$ observables $O_i$ ($i=1, ..N$), following the 
Bayesian approach\footnote{In particular, in the same spirit of section 
3 of \cite{Ciuchini:2000de}.}, the p.d.f. of the WCs can be written as
\begin{equation}
\rho \left( \mathbf{C}  \right) \propto \int
\prod^N_{i=1} \exp \left( - \frac{\left( O_i \left( \mathbf{C} ; 
\mathbf{x} \right) - \overline{O}_i \right)^2 }{2 \, \sigma^2_{O_i}} \right)
\prod_{j} f_{x_j}(x_j) d x_j \, \, .
\end{equation}
Let us now split the observables into two sets, $O^{+}$ and $O^{*}$ 
where the second (first) set refers to the observables (without) 
involving the decay $\BKstarmumu$. In a similar way, we split the set of 
inputs into three sets $\mathbf{x}= \{ \mathbf{x}^+, \mathbf{x}^*, 
\mathbf{x^c} \}$ where $\mathbf{x}^{+}$ ($\mathbf{x}^{*}$) are input 
parameters entering the set of observables $O^{+}$ ($O^{*}$) exclusively 
and $\mathbf{x^c}$ are the inputs which are common to both the sets of 
observables. The expression for the p.d.f. now reduces to
\begin{eqnarray}
\nonumber
\rho \left( \mathbf{C}  \right) & \propto  \int
\prod_{i} \exp \left( - \frac{\left( O^{+}_i \left( \mathbf{C} ; 
\, \mathbf{x}^{+}, \mathbf{x^c}  \right) -\overline{O^{+}_i} \right)^2} {2 \, 
\sigma^2_{O^{+}_i}} \right) \times
\prod_{j} \exp \left( - \frac{\left( O^{*}_j \left( \mathbf{C} ; 
\, \mathbf{x}^{*}, \mathbf{x^c}  \right) -\overline{O^{*}_j} \right)^2} {2 
\, \sigma^2_{O^{*}_j}} \right) \\
&\times \prod_{k_1} f_{x^{+}_{k_1}}(x^+_{k_1}) d x^+_{k_1} \times 
\prod_{k_2} f_{x^{*}_{k_2}}(x^*_{k_2}) d x^*_{k_2} \times 
\prod_{k_3} f_{x^{c}_{k_3}}(x^c_{k_3}) d x^c_{k_3} \, .
\end{eqnarray}
Now we would like to find the conditions such that the two sets of 
observables ``factorize''. In that case, we will be able to use the 
results of \cite{Descotes-Genon:2013wba} instead of redoing the full 
analysis. The factorization is possible if the common set of inputs 
$\mathbf{x^c}$ has small uncertainty compared to the other sources of 
uncertainties. If that is true then one can write to a very good 
approximation $f_{x^{c}_{k}}(x^c_{k})= \delta(x_k^c -\langle x_k^c 
\rangle)$ and the p.d.f. reduces to
\begin{equation}
\rho \left( \mathbf{C}  \right) \propto  \int
\prod_{i} \exp \left( - \frac{\left( O^{+}_i \left( \mathbf{C} ; 
\, \mathbf{x}^{+}, \mathbf{\langle x^c \rangle}  \right) - 
\overline{O^{+}_i} \right)^2} {2 \, \sigma^2_{O^{+}_i}} \right) \times
\rho_0 \left( \mathbf{C}  \right) \times \prod_{j} f_{x^{+}_{j}}(x^+_{j}) d x^+_{j} 
\end{equation}
where $\rho_0 \left( \mathbf{C} \right)$ is the result of a Bayesian 
analysis considering only the global analysis of the decay $B \to K^* 
\mu^+ \mu^-$. Using the result of \cite{Descotes-Genon:2013wba} in their 
section 3.2, we assume that
\begin{equation}
\rho_0 \left( C_7, C_9^{\mu} \right) \propto 
\exp \left( - \frac{\left( C_7 -\overline{C_7} \right)^2 }
{2\, \sigma^2_{C_7}} \right)
\exp \left( - \frac{\left( C^{\mu}_9 -\overline{C^{\mu}_9} \right)^2 }
{2 \, \sigma^2_{C^{\mu}_9}} \right) \, .
\end{equation}
with 
\begin{equation}
\overline{C_7} = -0.018 \, ,  \quad \sigma_{C_7}= 0.018 \, , \quad  
\overline{C^{\mu}_9} = -1.6 \, , \quad  \sigma_{C^{\mu}_9}= 0.3 \, .
\end{equation}
These values are inferred from eq.~(4) of \cite{Descotes-Genon:2013wba}. We 
finally remark that although Ref.~\cite{Descotes-Genon:2013wba} 
considered the observable $\mathcal{B} \left( B \to X_s \, \ell^+ \, 
\ell^- \right)$ under the assumption of lepton universality, we 
interpret that as $\mathcal{B} \left( B \to X_s \, \mu^+ \, \mu^- 
\right)$ and remove the experimental data on 
$\mathcal{B} \left( B \to X_s \, \mu^+ \, \mu^- \right)$ (our table \ref{tab-1}) 
from our fit for consistency. We have also not used the data on 
${\cal B}(\Bsmumu)$ since that was also already included in the fit of 
Ref.~\cite{Descotes-Genon:2013wba}.

\section{Input parameters}
\label{inputs}

\begin{table}[h!]
\begin{center}
\begin{tabular}{|l|lr|} 
\hline 
Parameters & \multicolumn{2}{c|}{Value} \\
\hline 
$G_F$                &  $1.166 \times 10^{-5}$\, GeV$^{-2}$  &  \cite{Beringer:1900zz}  \\
$\alpha_{em}(m_b)$   &  $1/133$                              &  \cite{Beringer:1900zz}   \\
$m_e$                &  $0.511$ MeV                          &  \cite{Beringer:1900zz} \\
$m_\mu$              &  $105.6$ MeV                          &  \cite{Beringer:1900zz}  \\
$m_b(m_b)$           &  $4.164$ GeV                          &  \cite{Khodjamirian:2010vf} \\
$m_{B^\pm}$          &  $5.279$ GeV                          &  \cite{Beringer:1900zz}  \\
$m_{K^\pm}$          &  $0.494$ GeV                          &  \cite{Beringer:1900zz}  \\
$\tau_{B^\pm}$       &  $1.641 \times 10^{-12}$ sec.         &  \cite{Beringer:1900zz}  \\ 
$|V_{ts}^* V_{ tb}|$ &  $40.58 \times 10^{-3}$               &  \cite{Charles:2004jd}   \\
$ m_{B^*_s} (1^-)$   &  $5.366$ GeV                          &  \cite{Beringer:1900zz} \\
\hline
\multicolumn{3}{|c|}{Wilson coefficients} \\
\hline
$C^{SM}_7(m_b)$ & -0.319 & \cite{Khodjamirian:2010vf} \\
$C^{SM}_9(m_b)$ & 4.228 & \cite{Khodjamirian:2010vf} \\
$C^{SM}_{10}(m_b)$ & -4.410 & \cite{Khodjamirian:2010vf} \\
\hline
\multicolumn{3}{|c|}{Form factors} \\
\hline
$f^+_{BK}$ & $0.34^{+0.05}_{-0.02}$ & \cite{Khodjamirian:2010vf} \\
$f^T_{BK}$ & $0.39^{+0.05}_{-0.03}$ & \cite{Khodjamirian:2010vf} \\
$b^+_1$ & $-2.1^{+0.9}_{-1.6}$ & \cite{Khodjamirian:2010vf} \\
$b^T_1$ & $-2.2^{+1.0}_{-2.0}$ & \cite{Khodjamirian:2010vf} \\
\hline
\end{tabular}
\caption{Input parameters.}
\label{tab:input} 
\end{center}
\end{table} 

\providecommand{\href}[2]{#2}\begingroup\raggedright\endgroup

\end{document}